\documentclass{aa}

\usepackage{psfig}

\newcommand{\lapeq}{\mbox{$~\stackrel{\scriptstyle <}{\scriptstyle \sim}~$}}
\newcommand{\gapeq}{\mbox{$~\stackrel{\scriptstyle >}{\scriptstyle \sim}~$}}

\def\n1808{NGC\,1808}
\def\m82{M\,82}
\newcommand{\hi}{H\,{\small{\sc I}}}
\newcommand{\hii}{H\,{\small{\sc II}}}

\begin{document}

%\thesaurus{to be filled in}

\title{A search for intergalactic H\,I gas in the NGC\,1808 
  group of galaxies}

\author{Michael Dahlem\inst{1}\thanks{\emph{Present address: 
Alonso de Cordova 3107, Vitacura, Casilla 19001, Santiago 19, 
Chile; email: mdahlem@eso.org}}
\and
Matthias Ehle\inst{2,3}
\and 
Stuart D. Ryder\inst{4}
}

\institute{Sterrewacht Leiden, Postbus 9513, 2300 RA Leiden, The
  Netherlands
\and
XMM-Newton Science Operations Centre, Apartado 50727, 28080 
  Madrid, Spain
\and
Astrophysics Division, Space Science Department of ESA, ESTEC,
  2200 AG Noordwijk, The Netherlands
\and
Anglo-Australian Observatory, P. O. Box 296, Epping, NSW 1710,
Australia
}

\offprints{M. Dahlem; mdahlem@eso.org}

\date{Received 21 March 2001 / Accepted 27 April 2001}

\abstract{
A mosaic of six \hi\ line observations with the Australia
Telescope Compact Array is used to search for intergalactic
gas in the NGC\,1808 group of galaxies. Within the field of
view of about $1\fdg4\times1\fdg2$ no emission from 
intergalactic \hi\ gas is detected, either in the form of 
tidal plumes or tails, intergalactic \hi\ clouds, or as gas 
associated with tidal dwarf galaxies, with a 5 $\sigma$ 
limiting sensitivity of about $3\,10^{18}$ cm$^{-2}$ (or 
$1.4\,10^7\ M_\odot$ at a distance of 10.9 Mpc, for the
given beam size of $127\farcs9\times77\farcs3$).
The \hi\ data of NGC\,1792 and NGC\,1808, with a velocity 
resolution of 6.6 km s$^{-1}$, confirm the results of
earlier VLA observations.
Simultaneous wide-band 1.34 GHz continuum observations
also corroborate the results of earlier studies. However,
the continuum flux of NGC\,1808 measured by us is almost
20\% higher than reported previously.
No radio continuum emission was detected from the type Ia
supernova SN1993af in the north-eastern spiral arm of 
NGC\,1808.
A comparison of NGC\,1792 and NGC\,1808 shows that it is not 
primarily the total energy input that makes the big difference 
between the starburst-related outflow in NGC\,1808 and the 
absence of such extraplanar features in NGC\,1792, but the 
area over which the energy released by stellar winds and 
supernovae is injected into the ISM.
}

\maketitle

\section{Introduction}

It has been argued by us earlier (Dahlem et al. 1990, 1994; 
Dahlem 1992, hereafter D92) that the starburst activity in the
S\' ersic-Pastoriza galaxy NGC\,1808 (S\'ersic \& Pastoriza
1965) and the high level of star formation (SF) in its nearby
companion galaxy NGC\,1792 might have been triggered by a 
gravitational interaction between these galaxies. This 
scenario had originally been proposed by van den Bergh (1978).
Indirect evidence comes, e.g., from a warp of the stellar and 
gaseous disk of NGC\,1808 and a pronounced asymmetry in the
spatial distribution of \hi\ gas and \hii\ regions in NGC\,1792. 
Except for the warp indicated above, which includes the stellar 
population in the outer ``$\theta$''-shaped spiral arms of 
NGC\,1808 (Koribalski et al. 1993; hereafter K93), the stellar 
distributions in both systems appear to be only slightly 
disturbed, arguing in favour of a distant passage. 
In particular the \hi\ gas in NGC\,1792 shows another peculiarity 
that made us speculate about the possibility of gas stripping 
in its outermost parts, namely a very notable sharp drop of the 
measured \hi\ surface brightness (and thus, under the assumption 
of optically thin emission, line of sight column density) {\it
within} the optically visible stellar disk (D92). 
This is unusual and normally found only in interacting systems,
like e.g. galaxies in the Virgo cluster (Kenney \& Young 1989). 
One galaxy detected by us exhibiting such behaviour {\it without} 
a nearby interaction partner is NGC\,3175 (Dahlem et al. 2001).
Most spiral galaxies, on the other hand, have detectable \hi\ 
gas far beyond their optically visible stellar disks (e.g. 
Bosma 1981). 

In a number of groups of galaxies intergalactic \hi\ has been 
found in the form of tails and/or plumes, as for example in
the NGC\,4631 group (Weliachew et al. 1978) and the Leo triplet,
which includes NGC\,3628 (Haynes et al. 1979). Such detections 
are important, because tidal features can provide strong 
constraints on n-body simulations of the dynamics of pairs
of interacting galaxies (see e.g. Combes 1978 for a model
describing the data by Weliachew et al. 1978). Such models 
help tremendously in establishing interaction timescales 
and understanding streaming motions of gas and stars in the 
individual systems, often providing the only way to explain 
the properties of the observed galaxies.
NGC\,1792 and NGC\,1808 belong to a small group of galaxies 
(Garcia 1993), for which we adopt here, as done by us 
previously, the distance of NGC\,1808 of $D = 10.9$ Mpc (based 
on $H_0 = 75$ km s$^{-1}$ Mpc$^{-1}$ and a virgocentric infall 
velocity of 300 km s$^{-1}$).
The disturbances mentioned above suggest a tidal interaction 
(e.g., Dahlem et al. 1994). However, the evidence for an interaction
in the past is circumstantial, which prompted us to search for 
intergalactic \hi\ gas in this group so as to obtain direct
proof of tidal forces at work. To that end we used the very 
compact 375-m array of the Australia Telescope Compact Array 
(ATCA; Frater et al. 1992)\footnote{The Australia Telescope is 
funded by the Commonwealth of Australia for operation as a 
National Facility managed by CSIRO.} to search for intergalactic
\hi\ gas in the NGC\,1808 group, conducting observations 
centred between NGC\,1792 and NGC\,1808, the two dominant
group members.
With a shortest baseline of 31\,m and good uv coverage on 
baselines up to 459\,m (discarding data from antenna 6, which 
is located at a distance of 3 km from the five moveable 
antennae), this array, being more compact than the VLA D 
array, is very well-suited to conduct such a search for 
extended low surface brightness emission.

\section{Observations and data reduction} 

\begin{table}[t!]
\begin{flushleft}
\leavevmode
\caption{Details of ATCA 375-m array observations}
\label{tab:obs}
\begin{tabular}{lc}
\noalign{\hrule\smallskip}
Observing Date & 1996, Dec. 8--9 \\
Net Observing Time & 11:00 h \\
Shortest Baseline & 31 m \\
Longest Baseline$^{\rm a}$ & 459 m \\
Flux Calibrator & 1934-638 \\
Phase Calibrator & 0537-441 \\
Primary Beam ({\it FWHM}) & $33'$ \\
%Radius of First Null & $..'$ \\
Restored Beam ({\it FWHM}) & $127\farcs9\times 77\farcs3$ \\
Bandwidth (helio. vel.) & 620--1675 km s$^{-1}$ \\
Velocity Resolution & 6.6 km s$^{-1}$ \\
Sensitivity per Channel$^{\rm b}$ & 5 mJy beam$^{-1}$ \\
$S_{\rm HI}/T_{\rm B}$ Conversion Factor$^{\rm c}$ & 62.2 Jy/K \\
Pointing\,1 ($\alpha,\delta(2000)$) 
  & 05:05:13.8, --37:58:49 \\
Pointing\,2 ($\alpha,\delta(2000)$) 
  & 05:05:40.9, --37:35:00 \\
Pointing\,3 ($\alpha,\delta(2000)$) 
  & 05:06:52.8, --37:40:06 \\
Pointing\,4 ($\alpha,\delta(2000)$) 
  & 05:06:52.9, --37:49:28 \\
Pointing\,5 ($\alpha,\delta(2000)$) 
  & 05:07:15.6, --37:54:33 \\
Pointing\,6 ($\alpha,\delta(2000)$) 
  & 05:07:42.1, --37:30:44 \\
\noalign{\smallskip\hrule}
\end{tabular}
\end{flushleft}
Notes to Table~\protect\ref{tab:obs}:\\
a) Discarding data from antenna 6. \\
b) 1-$\sigma$ rms at the centre of the field of view. \\
c) Conversion from antenna brightness temperature to flux density
   units for the angular resolution of our data of $127\farcs9\times 
   77\farcs3$ (ATCA Users Guide; October 1999). 
\end{table}

Our observations were carried out in late 1996; the most important
dates and parameters are listed in Table~\ref{tab:obs}. 
Of the 12:45 hours about 1:45 h were spent on flux, bandpass and
phase calibration. The rest of the time was split evenly among
six pointing directions. One was centred on NGC\,1792, one on
NGC\,1808 and the remaining four between the two galaxies, as 
specified in the table.

All data were calibrated in the usual way, using the MIRIAD
package installed at the Australia Telescope National Facility. 
We assumed a continuum flux density at 1.42 GHz of PKS\,B1934-638 
of 15.01 Jy. PKS\,B0537-441, which served as phase calibrator, was
observed at regular intervals during the observations.
The data for the six pointings were separated from each
other in the uv plane of the visibilities and calibrated 
separately.
Subsequently the data were recombined using the MIRIAD task 
UVAVER and then jointly cleaned, using the task MOSSDI, and 
restored 
%
% onto a $4''$ sampling grid 
%
with a common angular resolution (Full Width at Half Maximum) 
of {\it FWHM} = $127\farcs9\times77\farcs3$. 
For our attempt to detect extended low surface brightness 
emission, two original line channels each were averaged, 
resulting in a velocity resolution of 6.6 km s$^{-1}$.
The resulting data cube has 160 channels of 6.6 km s$^{-1}$
width each and covers a field of view (FOV) of approximately 
$1\fdg 4\times 1\fdg 2$. The major axis of this elongated FOV 
is oriented along the line connecting NGC\,1792 and NGC\,1808.
The data were also corrected for the effect of primary beam 
attenuation across the FOV. 

Continuum subtraction was performed by subtracting the average 
of the line-free channels from each of the channels containing 
line emission in each of the sub-datasets.

\section{Results and discussion}

\subsection{H\,I in NGC\,1792 and NGC\,1808}

\begin{figure*}
\psfig{figure=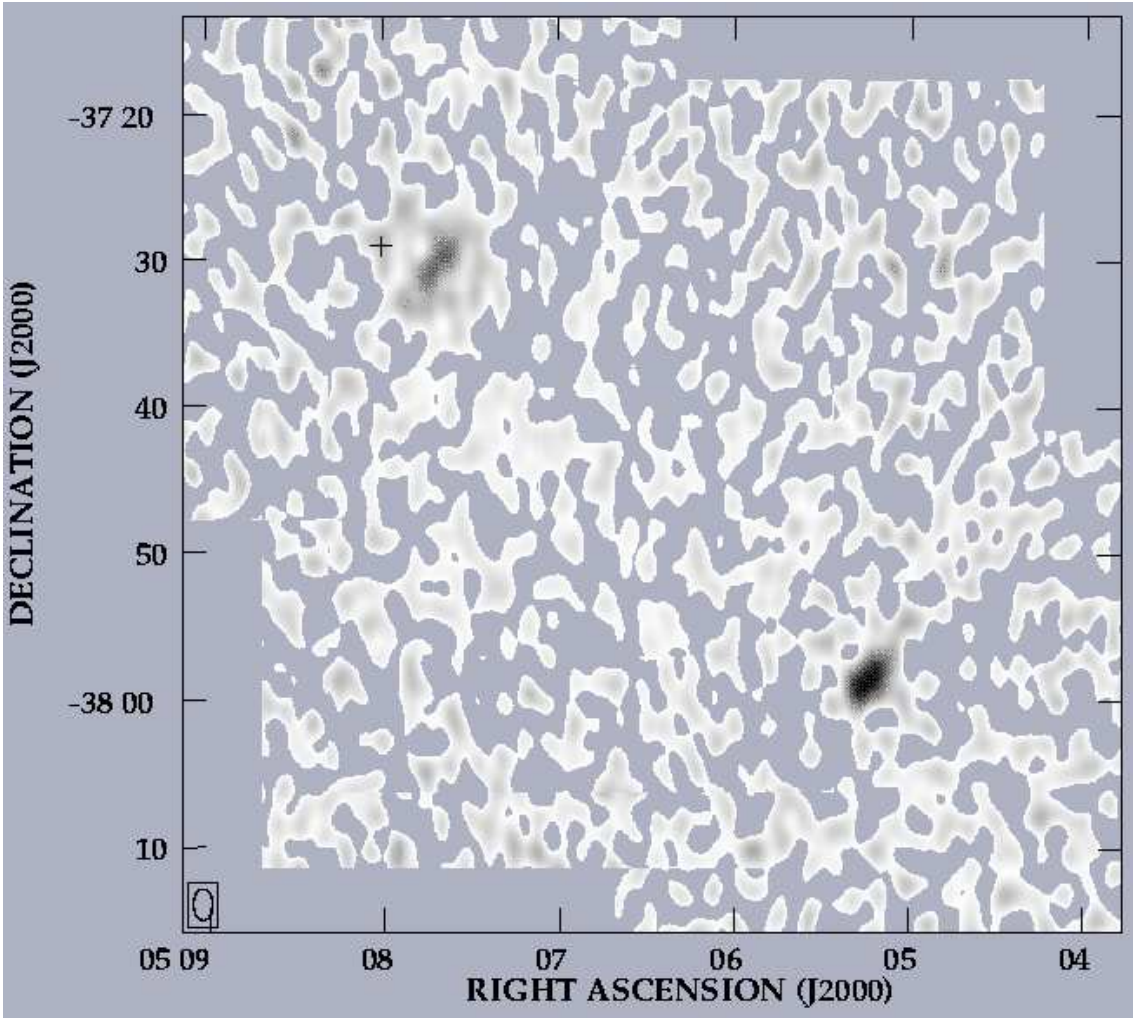,width=15.0cm}
\caption{
Image of the total \hi\ line emission of NGC\,1808 (upper left) 
and NGC\,1792 (lower right), the two dominant members of the 
NGC\,1808 group of galaxies, and their surroundings. The cross 
in the northeastern spiral arm of NGC\,1808 marks the position 
of SN1993af. The angular resolution of the data is indicated
in the lower left; the linear grey scale ranges from 0 to the 
maximum observed surface brightness of 1.912 Jy beam$^{-1}$.
\label{fig:hi}
} 
\end{figure*}

Our map of the integral \hi\ line emission from NGC\,1792, 
NGC\,1808 and the intergalactic space in between is displayed 
in Fig.~\ref{fig:hi}. 
This map was created using the AIPS routine MOMNT to 
calculate three ``moment maps'' (total intensity, velocity 
field and velocity dispersion) {\it from the original data 
cube},\footnote{Note that the ``standard'' procedure for the 
production of ``moment maps'' is to smooth the original
data cube and use a user-defined flux-density threshold in
that smoothed version as a ``mask'' to circumscribe features
that are considered to be real emission, while data points
below the threshold are blanked. In this process, the
signal-to-background ratio of the moment maps is increased
significantly. However, since features below a certain 
threshold are removed systematically, there is no measure 
any more of the inherent thermal noise of the data. Such 
blanked moment maps will be shown below. Here, in the 
display of the total FOV, we want to present an image 
including the true noise of the data, especially in the 
region of particular interest, between NGC\,1792 and 
NGC\,1808.}
before applying the primary beam correction.
It shows clearly resolved \hi\ emission from both galaxies. 
The 1-$\sigma$ rms noise of the data at the positions of both 
galaxies is of order 10 mJy beam$^{-1}$ per channel; at the 
centre of the FOV, where all six pointings overlap within the 
{\it FWHM} of the primary beam, it is approximately 5 mJy 
beam$^{-1}$ per channel.

Assuming optically thin \hi\ line emission ($\tau\ \ll\ 1$),
the observed flux densities can be converted into \hi\ column
densities following the relation (Eq.~12.2 by Giovanelli \&
Haynes 1988)

\begin{equation}
N({\rm H})\ = 1.83\,10^{18}\ T_{\rm B} \quad ,
\end{equation}

with the $T_{\rm B}/S_{\rm HI}$ conversion factor from 
Table~\ref{tab:obs}. $N$(H) is the \hi\ column density in units 
cm$^{-2}$ and $T_{\rm B}$ the antenna brightness temperature 
in units K. For our observations a line surface brightness of
1 Jy beam$^{-1}$ corresponds to a column density of $N({\rm H}) 
= 1.14\,10^{20}$ cm$^{-2}$ (assuming that the size of the 
observed structure is comparable to the beam size).

\begin{figure}
\hspace*{0.25cm}
\psfig{figure=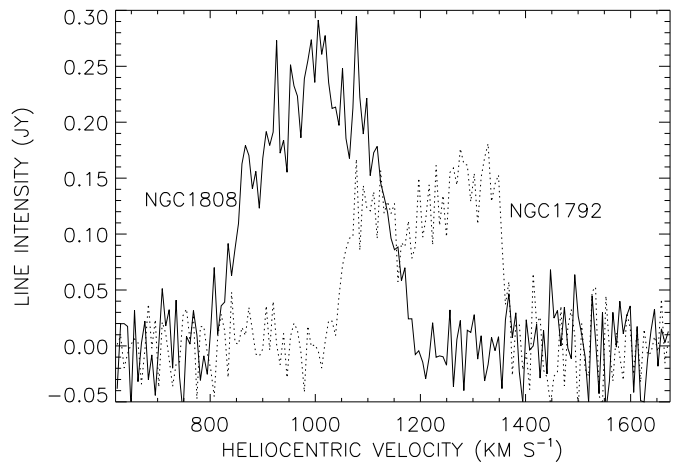,width=8.8cm}
\caption{Integral \hi\ line spectra of NGC\,1792 (dotted line)
and NGC\,1808 (solid line) superimposed onto each other to
illustrate the overlap in velocity space and the relative
intensities.
\label{fig:spec}
} 
\end{figure}

We will first introduce some global emission properties of 
both galaxies, before turning to details. 
The total \hi\ line spectra of both NGC\,1792 and NGC\,1808 
are displayed in Fig.~\ref{fig:spec}. The resolution of the 
current data, of 6.6 km s$^{-1}$, is about 3 times as high 
as that of earlier observations (D92; K93).
From these new spectra we derive the total widths of the integral 
\hi\ lines at the 20\%- and 50\%-level of the maximum, $W_{20}$
and $W_{50}$, respectively, the heliocentric systemic velocities, 
$v_{\rm sys}$, and the total \hi\ line fluxes, $f_{\rm HI}$. 
$f_{\rm HI}$ is calculated as the sum over all bins with
line emission from each galaxy, multiplied by the bin width
of 6.6 km s$^{-1}$. These values are listed in Table~\ref{tab:14}.

\begin{table}[t!]
\begin{flushleft}
\leavevmode
\caption{Integral properties of NGC\,1792 and NGC\,1808}
\label{tab:14}
\begin{tabular}{lrr}
\noalign{\hrule\smallskip}
Property & NGC\,1792 & NGC\,1808 \\
\noalign{\hrule\smallskip}
$W_{20}$ (km s$^{-1}$)        & $326\pm10$ & $342\pm10$ \\
$W_{50}$ (km s$^{-1}$)        & $295\pm10$ & $266\pm10$ \\
$v_{\rm hel}$ (km s$^{-1}$)   & $1207\pm10$ & $1001\pm10$ \\
$f_{\rm HI}$ (Jy km s$^{-1}$) & $39.5\pm5$ & $63.4\pm6$ \\
$f_{\rm 1.34}$ (mJy)          & $325\pm10$ & $616\pm15$ \\
log($L_{\rm 1.34}$) (W Hz$^{-1}$) & $21.66$ & $21.94$ \\
\noalign{\smallskip\hrule}
\end{tabular}
\end{flushleft}
%Notes to Table~\protect\ref{tab:14}: \\
\end{table}

All measurements of $f_{\rm HI}$, $W_{20}$ and $v_{\rm sys}$ 
confirm, within the error margins, earlier results obtained 
from VLA observations (D92; K93).
A few global properties derived from these global \hi\ data 
and other data are collated in Table~\ref{tab:basics}.

\begin{table*}[t!]
\begin{flushleft}
\leavevmode
\caption{Global properties$^{\rm a}$ of NGC\,1792 and NGC\,1808}
\label{tab:basics}
\begin{tabular}{lccc}
\noalign{\hrule\smallskip}
Property & Symbol & NGC\,1792 & NGC\,1808 \\
\noalign{\hrule\smallskip}
Absolute blue magnitude$^{\rm b}$   & $M_{\rm B}$ (mag)        
  & $-20.15\pm 0.10$ & $-20.24\pm 0.10$ \\
Total blue luminosity     & $L_{\rm B}$ ($L_\odot$)  
  & $1.6\,10^{10}$ & $1.7\,10^{10}$ \\
Total \hi\ gas mass       & $M_{\rm HI}$ ($M_\odot$) 
  & $1.1\,10^{9}$ & $1.8\,10^{9~~{\rm c}}$ \\
Total virial mass         & $M_{\rm T}$ ($M_\odot$)  
  & $0.54\,10^{11~~{\rm d}}$ & $1.1\,10^{11~~{\rm e}}$ \\
\hi\ gas to total mass ratio & $M_{\rm HI}/M_{\rm T}$ 
  & 0.021 & 0.016$^{{\rm c}}$ \\
\hi\ gas to blue light ratio & ${M_{\rm HI}\over L_{\rm B}}$ 
  (${M_\odot \over L_\odot}$) & 0.07 & 0.10$^{\rm c,e}$ \\
Total mass-to-light ratio & ${M_{\rm T}\over L_{\rm B}}$ (${M_\odot 
  \over L_\odot}$) & 3.4 & 6.5 \\
Optical colour index$^{\rm b}$ & $B-V$ (mag) & $0.68\pm 0.14$ & 
  $0.82\pm 0.14$ \\
Total FIR luminosity & $L_{\rm FIR}$ ($L_\odot$) 
  & $0.63\,10^{10~~{\rm d}}$ & $2.4\,10^{10~~{\rm f}}$ \\
Supernova Rate & $\nu_{\rm SN}$ (yr$^{-1}$) & 0.45 & 0.87 \\
Star Formation Rate$^{\rm g}$ & {\it SFR} ($M_\odot$ yr$^{-1}$) & 
  ~\,11.0 & 21.2 \\
Total Normalised {\it SFR} & ${{\it SFR}\over{A_{\rm SF}}}$
  (${{M_\odot}\over{{\rm yr}\ {\rm kpc}^{2}}}$) & 0.072 & 34.1$^{\rm h}$ \\
%
% distance modulus: -30.187
%
\noalign{\smallskip\hrule}
\end{tabular}
\end{flushleft}
Notes to Table~\protect\ref{tab:basics}: \\
a) All values without a note to a reference are calculated by us 
   based on $D = 10.9$ Mpc. \\
b) Both $B$ and $V$ magnitudes are from the RC3; the distance 
   modulus for $D = 10.9$ Mpc is --30.19. \\
c) Lower limit because of intrinsic self-absorption. \\
d) From Dahlem (1992). \\
e) From Koribalski et al. (1993). \\
f) From Dahlem et al. (1990). \\
g) For stars with $M \geq 5 M_\odot$. \\
h) Central starburst only; see text.
\end{table*}

Kinematic parameters, such as the kinematic position angle, 
{\it PA}, and the inclination angle, $i$, and their behaviour 
as a function of radial distance from the galaxy centre, can 
be best constrained from the earlier, higher-resolution VLA 
data.

\subsubsection{NGC\,1792}

\begin{figure*}
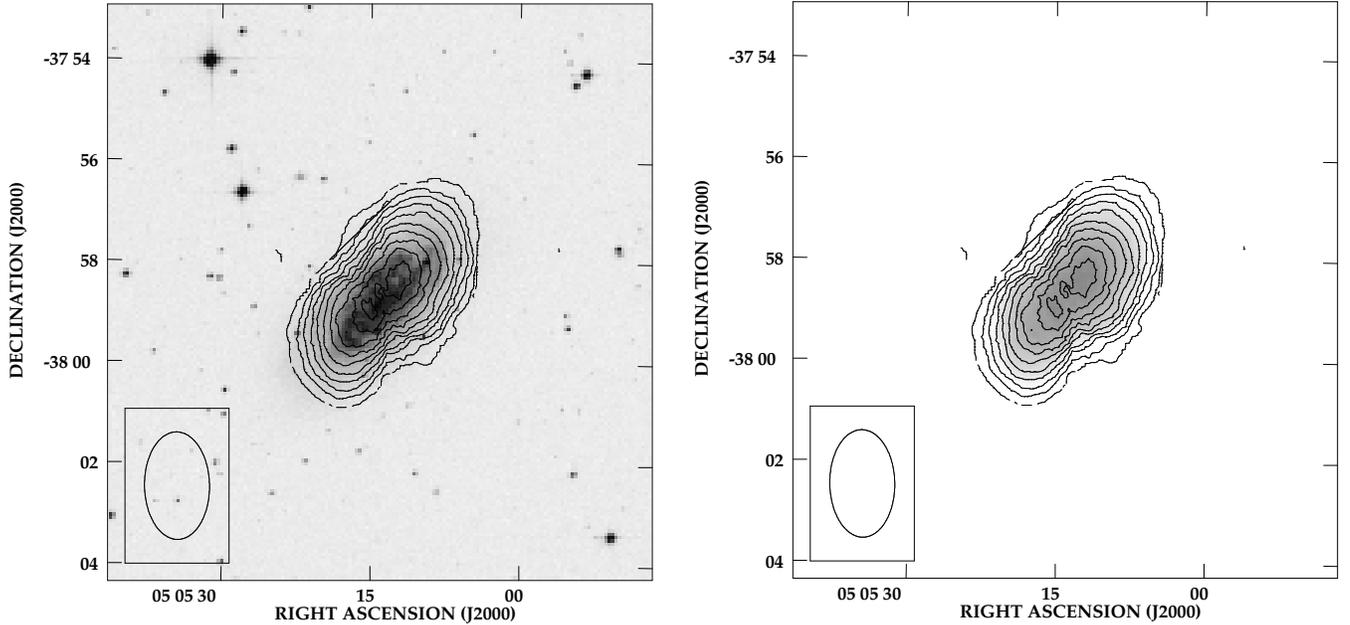

\psfig{figure=h2765f3a.ps,width=8.8cm}
\vspace*{-8.6cm}
\hspace*{9cm}
\psfig{figure=h2765f3b.ps,width=8.8cm}
\caption{
Contours of the total \hi\ line emission of NGC\,1792, overlaid 
on a DSS optical image (left panel) and contours with underlying
grey scale (right panel). The contour levels are 0.027, 0.11,
0.32, 0.54, 0.76, 0.97, 1.19, 1.41, 1.62, 1.84, 2.06\,$10^{20}$ 
cm$^{-2}$; the angular resolution is indicated in the lower left.
\label{fig:n1792hi}
} 
\end{figure*}

The \hi\ distribution in NGC\,1792 (Fig.~\ref{fig:n1792hi}) 
is compact. Not only is the extent of the \hi\ disk small 
compared to the stellar disk (D92), but with values up to
about $3\,10^{20}$ cm$^{-2}$ the \hi\ column densities 
are quite high (see Fig.~\ref{fig:hi}). Note that we use
here, contrary to Fig.~\ref{fig:hi}, the total intensity
map created by routine MOMNT, using a smoothed version of
the original datacube as a mask to avoid summing up noise
in areas where no line emission is present. This leads to
an improved signal-to-background ratio compared to the map 
in Fig.~\ref{fig:hi}\ (but leaves no sensible background
``noise'').
Even with its lower angular resolution our current map 
corroborates the steep outer gradient of the \hi\ emission 
distribution seen in the earlier VLA observations (D92), 
which is reminiscent of a sharp radial cutoff. 

Our data also confirm the regularity of the \hi\ velocity 
field of NGC\,1792 (not displayed).

\subsubsection{NGC\,1808}

\begin{figure*}
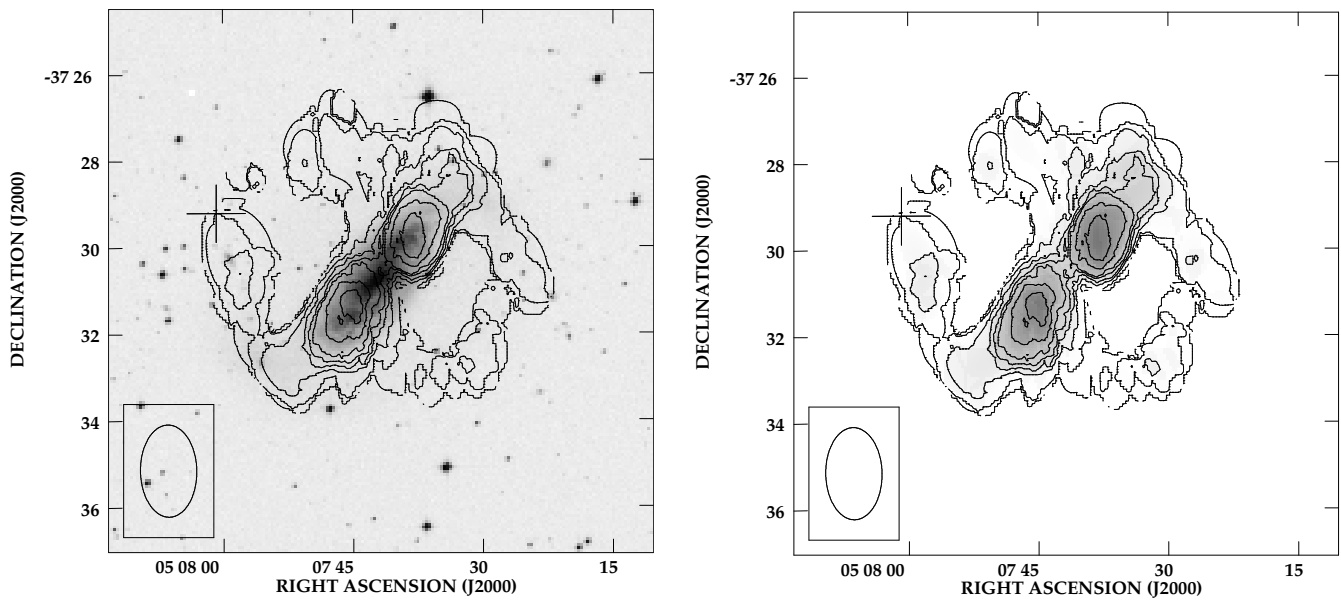

\psfig{figure=h2765f4a.ps,width=8.8cm}
\vspace*{-8.1cm}
\hspace*{9cm}
\psfig{figure=h2765f4b.ps,width=8.8cm}
\caption{
Contours of the total \hi\ line emission of NGC\,1808, overlaid 
on a DSS optical image (left panel) and contours with underlying
grey scale (right panel). The contour levels are 0.027, 0.054,
0.11, 0.22, 0.32, 0.43, 0.54, 0.76, 0.97\,$10^{20}$ cm$^{-2}$, 
the angular resolution is indicated in the lower left. The cross 
again marks the position of SN1993af.
\label{fig:n1808hi}
} 
\end{figure*}

Our \hi\ map of NGC\,1808 (Fig.~\ref{fig:n1808hi}) clearly 
resolves the inner disk and the outer spiral arms. 
Note that we use almost the same contour levels in 
Figs.~\ref{fig:n1792hi}\ and \ref{fig:n1808hi}, with only
a few more densely spaced low-level contours to display
the outer regions of NGC\,1808.
The \hi\ distribution is consistent with earlier results 
(e.g., Saikia et al. 1990; K93). The surface brightness and 
thus column density of the \hi\ emission from the inner disk 
is relatively low (about $0.2-1.0\,10^{20}$ cm$^{-2}$, 
typically). That of the outer spiral arms is yet lower, 
with values below $0.2\,10^{20}$ cm$^{-2}$, typically.
Part of the central \hi\ gas is not visible in emission 
because of intrinsic self-absorption along the line of sight 
towards the nuclear region (see K93 and Koribalski et al. 
[1996] for details). Therefore, the total \hi\ line flux
measured is a lower limit (Table~\ref{tab:14}). 
The cross marks the position of SN1993af, which will be 
discussed below.

The velocity field obtained from our data (not displayed) 
exhibits the same general features as observed in earlier 
VLA observations (K93), including the skewed appearance 
due to non-circular gas motions.

\subsubsection{A comparison of NGC\,1792 and NGC\,1808}

It is directly visible from Fig.~\ref{fig:hi}\ that the 
distribution of \hi\ in NGC\,1792 is much more compact than 
that of NGC\,1808. The extent of the \hi\ distribution in 
NGC\,1792 is smaller than that of its stellar disk and at 
the same time the surface brightnesses in the disk is also
higher. After correction for the different inclination
angles, the \hi\ column densities in NGC\,1792 are still
higher than in NGC\,1808, indicating that the mean gas 
density is higher.

These features of the \hi\ distributions in both galaxies 
can account for the observed distribution of SF activity: 
while NGC\,1808 has a circumnuclear starburst ring, 
NGC\,1792 exhibits widespread SF over large parts of its 
disk.
The low \hi\ column densities in the $\theta$-shaped outer 
spiral arms of NGC\,1808 suggest that there are no signs of 
SF there because of low gas densities.
A comparison of the SF properties of both galaxies will 
follow below, in the context of our radio continuum images 
(Sect.~\ref{par:rccomp}).

The rotational behaviour of both galaxies, NGC\,1792 and 
NGC\,1808, and the relative alignment of their spin vectors
was decribed by us in D92 (Fig.~5).

\subsection{No intergalactic H\,I between NGC\,1792 and 
  NGC\,1808}

Thus, the current observations corroborate earlier results
from dedicated studies of \hi\ in both NGC\,1792 and NGC\,1808. 
But, more importantly, they represent the up to now most 
sensitive search for intergalactic \hi\ gas in the vicinity 
of this pair of spirals.

The sensitivity of our data is illustrated by the detection 
of \hi\ in the outer spiral arms of NGC\,1808, despite the 
fact that the galaxy was observed in only part of the 
pointings. These spiral arms have relatively low column 
densities on the order of $10^{19}$ cm$^{-2}$. 
Near the centre of our FOV, the 1-$\sigma$ rms noise is 5 mJy 
beam$^{-1}$ per channel so that structures with column 
densities of $\gapeq 3\,10^{18}$ cm$^{-2}$ should have been 
detected at the 5-$\sigma$ confidence level. 
Based on a fiducial extent of roughly $5\times5$ kpc (one beam 
size) and a requirement to have a 5-$\sigma$ signal in three 
consecutive channels, this translates into a lowest detectable 
line flux of $f_{\rm H\,I}({\rm min}) \simeq 0.5$ Jy km s$^{-1}$. 
Using the relation by Roberts (1975),  

\begin{equation}
M_{\rm H\,I} = 2.356\,10^5\ D^2\ f_{\rm H\,I}\ [M_\odot] \quad ,
\end{equation}

where $D$ is the distance in units Mpc, and assuming optically 
thin emission, this corresponds to a minimum detectable \hi\ 
gas mass of approximately $1.4\,10^7\ M_\odot$ (or $5.6\,10^5\ 
M_\odot$ kpc$^{-2}$) at the distance of the NGC\,1808 group. 
Near the boundaries of the FOV, where both NGC\,1792 and NGC\,1808
are located, these values increase to approximately $3\,10^7\ 
M_\odot$ (or $1.2\,10^6\ M_\odot$ kpc$^{-2}$).
The velocity range covered by our observations is 1055 km 
s$^{-1}$, from 620 to 1675 km s$^{-1}$ (Table~\ref{tab:obs}), 
centred at 1148 km s$^{-1}$, i.e. between the systemic
velocities of NGC\,1792 and NGC\,1808. This makes it 
unlikely that any gas between NGC\,1792 and NGC\,1808 was 
missed in redshift space.

We used not only the data with the highest possible angular
and velocity resolution, but also various smoothed versions 
of the data cube. No significant \hi\ emission was detected 
that is not displayed in Figs.~\ref{fig:hi}, \ref{fig:n1792hi}\
and \ref{fig:n1808hi}.

It is evident from Fig.~\ref{fig:hi}\ that, despite the use 
of a very compact interferometer configuration, no \hi\ gas 
could be detected in the space between NGC\,1792 and NGC\,1808 
and in their immediate surroundings, at the sensitivity of 
the data. Only small amounts of diffuse \hi\ gas might have 
eluded detection, as quantified above.
This is not entirely unexpected because, with the gas-to-total 
mass ratios listed in Table~\ref{tab:basics}, neither of the 
two galaxies is prominently \hi -deficient when compared with 
the large sample used by Roberts \& Haynes (1994) to determine 
the ``typical'' properties of galaxies of different Hubble 
types.

\subsubsection{Absence of tidal features}

At the boundaries of the \hi\ emission distributions of 
both galaxies, NGC\,1792 and NGC\,1808, no plumes, tails or
other previously undetected tidal features were found at a 
significant level in addition to the \hi\ emission features 
already known from dedicated observations centred on the 
individual galaxies, with a detection limit of order 
$3\,10^7\ M_\odot$.

These results imply that the interaction of the pair NGC\,1792
and NGC\,1808 that might have caused the observed disturbances
(D92; K93) {\it must} have left the galaxies with most of their
gas. Therefore, a far-field encounter is still the most likely
explanation of all observations of these systems.

Our results leave no doubt that the NGC\,1808 group does 
not lend itself to dynamical studies by means of numerical 
modeling, because there are no distinct features that could 
sensibly constrain the boundary conditions for n-body 
simulations to model a putative interaction.

\subsubsection{No dwarf galaxies detected}

It is also interesting to note that our observations have not
revealed any signs of \hi\ emission from previously unknown 
dwarf galaxies near NGC\,1792 or NGC\,1808, as found in other 
groups like for example the NGC\,4666 group of galaxies 
(Walter et al., in preparation), see also e.g. Brinks (1990).

This again implies that there are no significant amounts of 
debris in the form of tidal dwarf galaxies in the NGC\,1808 
group similar to those found for the first time at the end 
of the tidal arms of the ``Antennae'' galaxies (NGC\,4038/39, 
Mirabel et al. 1992) and later e.g. in the Leo Triplet of 
galaxies (Chromey et al. 1998).

\subsubsection{No primordial intergalactic H\,I clouds}

Our results also bear some significance in the context of the
hypothesis originally proposed by Oort (1966) and recently
refined by Blitz et al. (1999) that Galactic \hi\ high-velocity 
clouds (HVCs) might be primordial \hi\ gas clouds raining
down on the Galaxy. Assuming that similar \hi\ gas clouds
exist in external groups of galaxies, these should be 
detectable with the sensitivity of our observations.

Although our data do not span the entire NGC\,1808 group of
galaxies, the non-detection of intergalactic \hi\ at the 
sensitivity stated above implies a constraint on the possible
existence of intergalactic primordial \hi\ clouds.
No clouds are detected down to a mass of about $10^7\ M_\odot$.
This implies that such clouds, if existent, cannot form
a significant fraction of the total \hi\ gas mass of that part 
of the NGC\,1808 group of galaxies around the two spirals 
NGC\,1792 and NGC\,1808 sampled by our observations.
This, in turn, constrains the role that such primordial gas
can play in forming HVCs raining down on the two galaxies
NGC\,1792 and NGC\,1808, following the reasoning by Zwaan
(2001). In the NGC\,1792/NGC\,1808 system intragroup \hi\
gas clouds cannot play a significant role as the potential
progenitors of HVCs. This implies that our observations cannot
lend any support to the theory by Oort (1966) and Blitz et al.
(1999).
Detailed calculations and simulations based on \hi\ observations 
of five groups of galaxies are given by Zwaan (2001) and in the 
references therein.

\subsubsection{No intergalactic medium at all?}

This non-detection of \hi\ emission on different spatial scales 
does {\it not} exclude the possibility that an intergalactic 
medium might exist. 
However, our data put very tight constraints on the properties
of such a medium. If located between NGC\,1792 and NGC\,1808
and if in the form of \hi, this gas must have very low column 
densities ($\lapeq\ 3\,10^{18}$ cm$^{-2}$).
There might also be \hi\ gas outside the field of view of our 
observations, farther away from the two galaxies, towards the
other group members (Garcia 1993). Although these are at
yet larger projected distances from NGC\,1792 and NGC\,1808
and/or have lower total masses, it is conceivable that other
members of the group, such as e.g. the late-type system
ESO\,305-G009 at a recession velocity of 1025 km s$^{-1}$, 
might have taken part in a past interaction.

One might hypothesise whether any intergalactic gas might
exist in the form of other phases. For example, at least
part of the gas could possibly be too warm to be visible in 
\hi\ emission. 
Donahue et al. (1995) detected a diffuse H$\alpha$-emitting 
medium around NGC\,4631, which might also be related to the 
softest observed X-ray emission at k$T \simeq 0.05$ keV 
(Dahlem et al. 1998). 
Searches which could detect warm, H$\alpha$-emitting gas in 
the NGC\,1808 group of galaxies have not yet been conducted. 
But it is unlikely that such gas exists, because warm ionised
gas should normally be accompanied by a neutral component,
as in the case of the NGC\,4631 group, which we did not 
detect around NGC\,1792/1808.
Both the sensitivity and the field of view of the ROSAT 
observations of NGC\,1792 and NGC\,1808 (Dahlem et al. 1994; 
Junkes et al. 1995) should have led to the detection of a hot 
($\sim 10^6$ K) intergalactic medium, if such should exist. 
However, no hot ionised gas was found.

At the other end of the energy scale, the gas might be too 
cold to be seen in emission. Our \hi\ emission observations
trace only {\it warm} \hi\ gas. Cold neutral hydrogen gas 
would not show up in our data, being only observable in 
absorption against background sources. The gas might be yet
colder and exist in the form of molecular gas. However, such 
gas, outside self-shielding gas clouds and in the absence
of significant amounts of shielding dust in galaxy disks, 
is exposed to the intergalactic UV radiation field and thus
expected to be heated to temperatures that would make it 
visible in emission (Shull et al. 1999). 

A search for signatures of stellar plumes or tails on deep 
optical images (see Chromey et al. 1998 for an example)  
in member galaxies of the NGC\,1808 group of galaxies,
including ESO\,305-G009, also led to a negative result 
(D. Malin; private communication).

Although being far from conclusive, all these arguments 
make it very unlikely that there are significant amounts 
of intergalactic gas in the NGC\,1808 group of galaxies, at 
least in the form of gas phases that are presently known to 
us (and observable).

\subsection{1.34 GHz continuum imagery}

\begin{figure*}
\psfig{figure=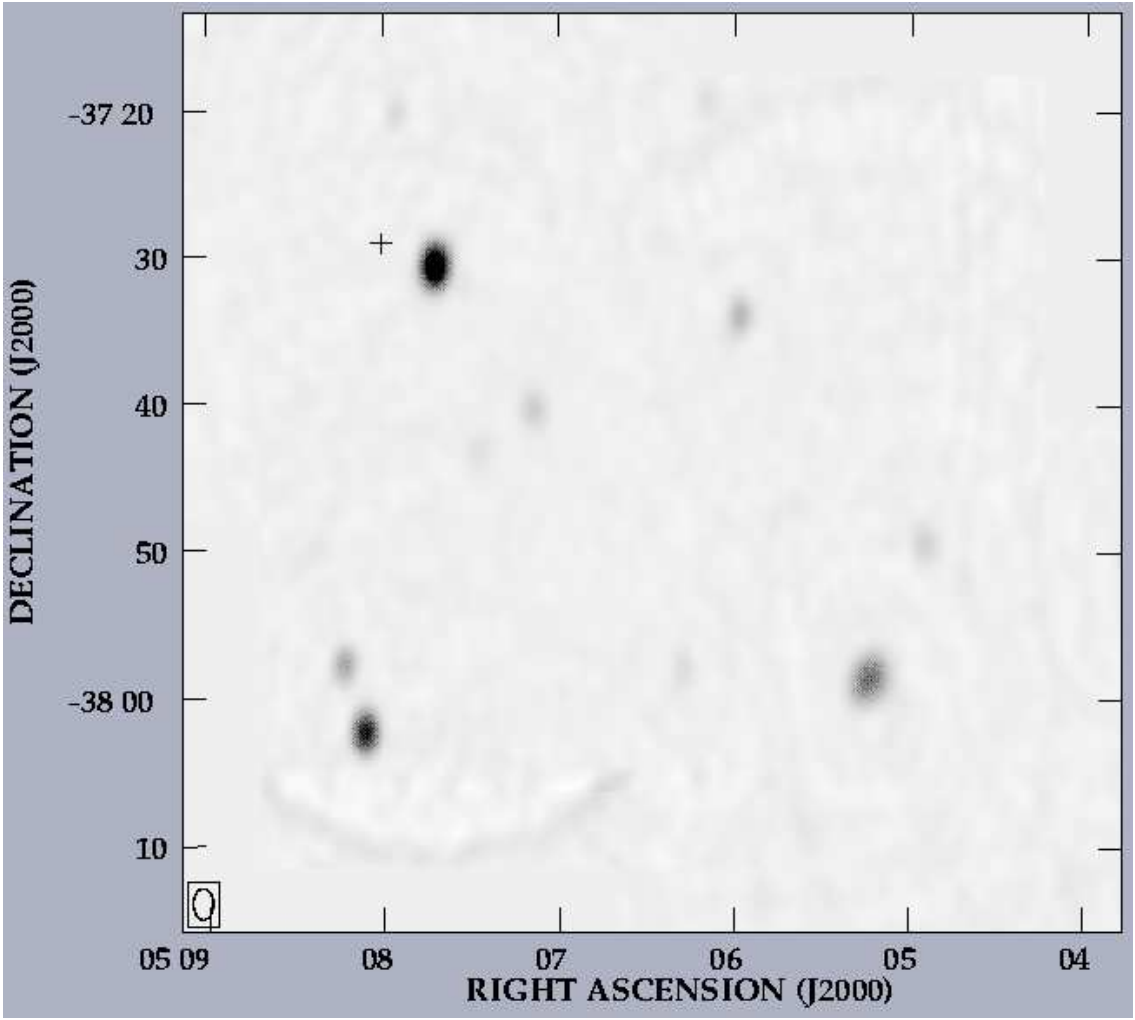,width=15.0cm}
\caption{
Image of the 1.34 GHz radio continuum emission of NGC\,1808 
(upper left) and NGC\,1792 (lower right), the two dominant 
members of the NGC\,1808 group of galaxies, and their 
surroundings. The cross in the northeastern spiral arm of 
NGC\,1808 marks the position of SN1993af. The angular
resolution of the data is indicated in the lower left; the
linear grey scale ranges from 0 to the maximum observed surface 
brightness of 484 mJy beam$^{-1}$. The rms noise is about
1 mJy beam$^{-1}$ near the centre and 2 mJy beam$^{-1}$
in the vicinity of SN1993af.
\label{fig:rc}
} 
\end{figure*}

The setup of the ATCA allows for parallel \hi\ observations
at one intermediate frequency (IF) and continuum observations
at the second IF. Therefore, 1.34 GHz radio continuum images 
with a bandwidth of 84 MHz were obtained simultaneously, 
which we can compare with the existing data in the literature 
(cf. Table~\ref{tab:14}). The map of the same field of view
as the \hi\ total intensity image (Fig.~\ref{fig:hi}) is 
displayed in Fig.~\ref{fig:rc}. NGC\,1792 and NGC\,1808 are
the most prominent radio continuum sources in the field of
view. In the lower left an arc of an imperfectly cleaned
beam fringe of NGC\,1808 is visible. The rms noise of the
map is about 1 mJy beam$^{-1}$ in the central region and
about 2 mJy beam$^{-1}$ further out, near NGC\,1792 and
NGC 1808.

\subsubsection{NGC\,1792}

The 1.34 GHz continuum emission distribution of NGC\,1792
derived from the present data is consistent with our earlier 
map (D92). 
The total flux measured from the present data is $325\pm10$ 
mJy. This is compatible with our earlier measurement of 
$324\pm20$ mJy at 1.41 GHz (D92), which implies that the 
flux calibration of our observations is consistent with the 
earlier VLA observations.

The radio continuum surface brightness is high, which is 
consistent with the high \hi\ surface brightnesses (D92
and above) and the ``extremely high-surface-brightness 
arms'' visible in optical images (Sandage \& Brucato 1979).

\subsubsection{NGC\,1808}

The 1.34 GHz continuum emission of NGC\,1808 is almost
unresolved by our observations. The resulting image does
not add new information to previously published results
(Dahlem et al. 1990; Saikia et al. 1990). 
The total flux derived from the present data is $616\pm15$ 
mJy, which is about 18\% higher than our earlier result of 
$523\pm10$ mJy at 1.49 GHz (Dahlem et al. 1990). Partly, 
this might be explained by the different observing 
frequencies, but part of the discrepancy might be due
to an intrinsic variability, possibly of the low-luminosity
AGN (V\'eron-Cetty \& Veron 1985; Awaki \& Koyama 1993;
Awaki et al. 1996).

We checked our data for continuum emission from the type 
Ia supernova SN1993af at $\alpha,\delta(2000) = 05^{\rm h}\ 
08^{\rm m}\ 00\fs79, -37^\circ\ 29'\ 12\farcs8$ (Hamuy and 
Maza 1993; Biegelmayer et al. 1993), in the north-eastern 
outer spiral arm of NGC\,1808. We have not detected any 
significant radio continuum emission at the position of 
the supernova, with a 1-$\sigma$ noise level of our map 
in this region of about 2.5 mJy beam$^{-1}$.

\subsubsection{Comparing the radio continuum properties}
\label{par:rccomp}

Considering that NGC\,1808 is a classical starburst galaxy,
while its companion NGC\,1792 is a ``normal'' spiral, it
is noteworthy that NGC\,1792's total radio continuum flux
density at 1.34 GHz is more than half that of NGC\,1808 
(Table~\ref{tab:basics}; see also Junkes et al. 1995).
The major difference between the two galaxies does not lie
in the total energy output, but in the size of the region 
over which the energy is produced in stellar winds and 
supernovae (Leitherer \& Heckman 1995) and released into
the ISM.

In NGC\,1808, about 90\% of the radio continuum emission
arises from the central starburst, i.e. inside a region 
of $16''\times10''$ (Saikia et al. 1990). Assuming an
inclination angle of about $60^\circ$ and thus circular 
symmetry of the starburst, with a major axis extent of
$16''$, this corresponds to an area of about 0.56 kpc$^2$. 
Based on a total radio continuum flux of 616 mJy at 1.34 
GHz, this implies that about 555 mJy arise from the central 
region. 

At the distance of NGC\,1808 of 10.9 Mpc, this corresponds
to a radio luminosity of $L_{1.34}\ = 8.75\times10^{20}$ W
Hz$^{-1}$. 
Using equations 18, 20 and 21 by Condon (1992), and assuming
that practically all radio continuum emission at 1.34 GHz
is synchrotron radiation, this radio luminosity can directly 
be translated into a supernova rate, $\nu_{\rm SN}$, and a 
star formation rate for stars with $M\ \geq\ 5\ M_\odot$, 
{\it SFR} (see Table~\ref{tab:basics}). 
For NGC\,1808 a mean radio spectral index of $\bar\alpha = 
-0.88$ ($S\ \propto\ \nu^\alpha$) was adopted (Dahlem et 
al. 1990).
Strictly speaking, this value is valid only for the central 
starburst, but since the radio emission of the starburst 
dominates the total flux, it should be a good approximation
of the overall spectral index.
From this, it follows that the normalised {\it SFR} {\it within 
the circumnuclear starburst} in NGC\,1808 (based on 90\% 
of the total nonthermal radio continuum flux density) is 
approximately 34.1\,$M_\odot$ yr$^{-1}$ kpc$^{-2}$.

In NGC\,1792 the SF activity is much more wide-spread, out 
to a radius of $2\farcm2$ (7.0 kpc; D92), leading to an 
area over which SF is observed, again under the assumption 
of circular symmetry, of 153.1 kpc$^2$. 
Based on a radio spectral index of $\bar\alpha = -0.82$ (D92), 
the total {\it SFR} following from the observed 1.34 GHz flux 
density of 325 mJy is 11.0\,$M_\odot$ yr$^{-1}$, which then 
translates into a normalised {\it SFR} per unit area in NGC\,1792 
of about 0.072\,$M_\odot$ yr$^{-1}$ kpc$^{-2}$.

Thus, the primary reason for the different excitation conditions 
of the ISM in NGC\,1792 and NGC\,1808 and the prominent outflow 
from the central starburst region of NGC\,1808, which has no 
equivalent in NGC\,1792, is the difference in area over which 
the energy produced by stellar winds and supernovae is released, 
as reflected by the observed difference in radio continuum surface 
brightnesses in Fig.~\ref{fig:rc}. This dilution effect alone 
leads to a difference in the average energy density by a factor 
of roughly 270, while the total difference in the normalised {\it 
SFR} between the disk of NGC\,1792 and the central starburst in 
NGC\,1808 is a factor of 474.

For more detailed information on the SF activity in both
galaxies see e.g. Saikia et al. (1990), Forbes et al. (1992), 
Phillips (1993), Collison et al. (1994), Dahlem et al. (1994)
Krabbe et al. (1994), Mazzolini \& Webster (1996) and Kotilainen 
et al. (1996).
The high-resolution rotation pattern of \hi\ gas seen in 
absorption against the starburst in NGC\,1808 is described 
by Koribalski et al. (1996).
The distribution of stars in the central $2'\times2'$ of 
both galaxies can be seen in the NIR $H$ band images by
Jungwiert et al. (1997).

\vspace{0.5cm}

\noindent{\it Acknowledgements:} We would like to thank L. 
Staveley-Smith for his help in the calculation of the flux 
density to column density conversion and D. Malin for 
communicating the absence of intergalactic features on his 
unpublished deep optical images of the NGC\,1808 group of 
galaxies.
The Digitized Sky Surveys were produced at the Space Telescope
Science Institute under U.S. Government grant NAG W-2166. The
images of these surveys are based on photographic data obtained 
using the Oschin Schmidt Telescope on Palomar Mountain and the 
UK Schmidt Telescope. The plates were processed into the present
compressed digital form with the permission of these institutions. 

%\appendix

%\clearpage

\end{document}